# Anisotropic Electrical Spin Injection in Ferromagnetic Semiconductor Heterostructures

D.K. Young, E. Johnston-Halperin, and D.D. Awschalom [a]

*Center for Spintronics and Quantum Computation,
University of California, Santa Barbara, CA 93106, USA*

Y. Ohno and H. Ohno

*Laboratory for Electronic Intelligent Systems, Research Institute of Electrical Communication,
Tohoku University, Katahira 2-1-1, Aoba-ku, Sendai 980-8577, Japan*

Abstract

A fourteen-fold anisotropy in the spin transport efficiency parallel and perpendicular to the charge transport is observed in a vertically-biased (Ga,Mn)As-based spin-polarized light emitting diode. The spin polarization is determined by measuring the polarization of electroluminescence from an (In,Ga)As quantum well placed a distance $d$ (20–420 nm) below the p-type ferromagnetic (Ga,Mn)As contact. In addition, a monotonic increase (from 0.5 to 7%) in the polarization is measured as $d$ decreases for collection parallel to the growth direction, while the in-plane polarization from the perpendicular direction (~0.5%) remains unchanged.

Understanding the physical mechanisms underlying the manipulation of electronic spin in semiconductors may ultimately lead to multifunctional devices based on photonics, electronics, and magnetics.[1] Moreover, utilizing coherent spin phenomena in semiconductors[2] may be fundamental for the future of quantum computation in the solid state. The demonstrations of electrical spin injection into semiconductors using both ferromagnetic[3] and paramagnetic semiconductors[4], and more recently with a ferromagnetic metal[5] and Zener tunneling processes[6,7] are promising for potential spin based electronics.

Here we report a fourteen-fold anisotropy in the electrical spin injection efficiency between directions parallel and perpendicular to the current flow along the growth axis in a spin-polarized light emitting diode (LED),[3] demonstrating the importance of device geometry in obtaining efficient injection and detection. Under forward bias, spin-polarized holes[8,9] from (Ga,Mn)As and unpolarized electrons from an n-type GaAs substrate are injected into an embedded (In,Ga)As quantum well (QW) separated from the ferromagnetic region by a spacer layer $d$ varying from 20–420 nm in thickness. Spin polarization of the electrically injected holes is measured by analyzing the polarization (P) of the emitted electroluminescence (EL) either along the growth direction (through the substrate) or in plane (from a cleaved facet). In addition, we find that as the spacer layer thickness decreases, the magnitude of EL polarization monotonically increases from 0.5 to 7% when the hole spin orientation is along the direction of charge transport (growth direction). In contrast, EL polarization is insensitive to spacer layer thickness when measured in the plane of the sample (P ~ 0.5% for all $d$), where the hole spin orientation is perpendicular to the charge transport. This spacer layer dependence is not intrinsic to the QW, but arises from a difference in spin transport efficiency for the two geometries.

The device structure shown in the inset of Fig. 1(a) is grown by molecular beam epitaxy on a (100) n-GaAs substrate with a 500 nm $n^+$-GaAs buffer layer (doping density $N_D = 2 \times 10^{18}$ cm$^{-3}$) and the following layers: 20 nm undoped GaAs, 10 nm undoped In$_{0.12}$Ga$_{0.88}$As strained QW, undoped GaAs spacer with thickness $d$ (20, 70, 120, 220 or 420 nm), and 300 nm Ga$_{1-x}$Mn$_x$As with $x$ = 0.045 or 0.035. Details of the growth of the magnetic layer can be found elsewhere.[10] The epitaxial wafer is processed into light emitting devices having 150 μm-wide mesa stripes defined by wet chemical etching after metal electrode deposition (5 nm Ti/250 nm Au) and cleaved into ~1 mm x 5 mm pieces. Both p and n contacts are made from the top allowing EL collection from a cleaved facet or through the substrate [Fig. 1(a) inset]. Two sets of control samples are prepared to verify spin injection, 1) a nonmagnetic device ($d$ = 20 nm) with a p-type GaAs:Be layer ($p = 2 \times 10^{18}$ cm$^{-3}$) substituted for the (Ga,Mn)As layer and 2) a magnetic structure ($d$ = 100 nm) without metal contacts enabling resonant optical excitation of the QW.

The spontaneous magnetic ordering below the Curie temperature ($T_C$) in (Ga,Mn)As results in a spin-polarized hole gas.[9] Under forward bias conditions, these spin-polarized holes are injected into the QW through the undoped GaAs spacer layer, while unpolarized electrons are supplied from the n-GaAs substrate. The samples are mounted in a magneto-optical cryostat with a variable magnetic field, applied in or out of plane that is monitored by *in-situ* Hall bars. For both cases, EL is collected along the applied field axis. The polarization P = (I$^+$ - I$^-$)/(I$^+$ + I$^-$) of the EL spectra is analyzed with a variable wave plate and linear polarizer, and is detected with a charge coupled device attached to a 1.33 m spectrometer. Here I$^+$ and I$^-$ are intensities of the right and left circularly polarized components of the EL, respectively.

Figures 1(a-c) show the optical and electrical characteristics at T = 5 K for a device with $d$ = 70 nm. Figure 1(a) shows the EL intensity as a function of energy for different bias conditions and its I–V curve is shown in Fig. 1(b). Moreover, the (In,Ga)As QW emission is spectrally distinct (E = 1.39 eV) from that of the GaAs heterostructure (E = 1.51 eV) allowing one to study the depth of spin injection with varying spacer layer.[3,4] Figure 1(c) shows the polarization (●) and EL intensity (solid curve) as a function of energy with an out of plane magnetic field $H_\perp$ (~5 kOe). Peaks in the EL intensity (FWHM = 10 meV) and polarization coincide with the QW ground-state emission energy indicating that spin-polarized holes are injected into the QW. We observe minimal dependence of the polarization on the injection current density,[3] allowing us to drive the device for optimal signal to noise. Finally, we characterize the magnetization of the (Ga,Mn)As layer at T = 5 K by superconducting quantum interference device (SQUID) magnetometry [Fig. 1(d)] confirming that easy and hard magnetization axes of the (Ga,Mn)As layer are in and out of the sample plane, respectively.[11]

Figure 2(a) shows relative changes in EL polarization[12] $\Delta P = P - P_{background}$, as a function of magnetic field ($H_\perp$) for various temperatures near and below $T_C$. Below T = 62 K, $\Delta P$ saturates around $H_\perp$ ~ 2.5 kOe, tracking the magnetization of the (Ga,Mn)As shown in Fig. 1(d). The saturation polarization $P_S$ decreases and ultimately vanishes as the temperature increases from T = 5 to 62 K, commensurate with the temperature dependent magnetization measured by the SQUID (inset). The deviation from mean field theory of M(T) is consistent with previous studies.[8,9,11]

The non-magnetic device (d = 20 nm) is measured in order to verify that the field dependence of the polarization is due to spin injection rather than Zeeman splitting induced by stray fields from the (Ga,Mn)As. In contrast to the magnetic devices, the EL polarization from the non-magnetic device [Fig. 2(b)] does not show saturating behavior as a function of field, revealing only the Zeeman contributions in the QW polarization[12] (P = 0.5% at $H_\perp$ = 5 kOe). This indicates that Zeeman splitting in the QW from the applied field as well as the local fields from the (Ga,Mn)As layer ($H_{stray}$ ~ 500 Oe)[10] are unlikely to be responsible for the saturating polarization in the magnetic structures.

Since (Ga,Mn)As exhibits strong magnetic circular dichroism (MCD),[9] it is also important to confirm that the observed saturating polarization is not due to preferential re-absorption of QW luminescence passing through the (Ga,Mn)As layer. A magnetic sample without metal contacts is prepared, allowing resonant optical excitation of unpolarized carriers into the QW in the same measurement geometry as used for the EL. A *p*-type layer between the QW and a semi-insulating substrate is incorporated into the structure in order to reduce the electrostatic potential across the junction, thus leading to more efficient radiative recombination. A pulsed Ti-sapphire laser (FWHM ~20 meV) is used to create *unpolarized* carriers in the QW by illuminating through the cleaved facet with linearly polarized light at $E = 1.401$ eV, 56 meV above the QW ground state, and ~100 meV below the GaAs band gap. The photoluminescence polarization as a function of magnetic field shown in Fig. 2(c) reveals no saturation, suggesting that the EL polarization does not originate from MCD effects.

Optical selection rules responsible for the QW luminescence[13] suggest that the measured spin polarization depends on collection geometry. By rotating the sample 90°, we measure from the cleaved edge, and observe hysteretic EL polarization [shown in Fig. 2(d)], reflecting the in-plane magnetic properties of the (Ga,Mn)As layer[3] [Fig. 1(d)]. However, the spin polarization is a factor of 10 smaller than out of plane, and exhibits an overall minus sign in the field dependence. Due to quantum confinement and strain, angular momentum of the heavy hole (HH) is pinned along the growth direction, and in plane for the light hole (LH).[13] Therefore, non-zero polarization from both in and out of plane geometries suggests a contribution from both spin-polarized heavy and light holes to the EL. Similar behavior has also been observed in spin polarized Zener tunneling diodes[6] as well as spin ejection studies.[14]

In an attempt to determine whether the polarization anisotropy depends on a difference in spin transport efficiency or is an intrinsic property of the QW, a set of samples with varying spacer layer thickness ($d = 20$–420 nm) was studied [shown in Fig. 3(a)]. As the magnetic layer is placed closer to the QW, the magnitude of the EL saturation polarization $\Delta P_S$ increases from 0.5 to 7% when the hole spin is oriented along the charge transport direction (growth direction).

In contrast, when the hole spin is oriented orthogonal to charge transport the magnitude of the remanent EL polarization remains constant ($\Delta P_R \sim 0.5\%$) over the range of spacer layer thicknesses (inset), consistent with earlier measurements.[3] If the factor of fourteen enhancement was intrinsic to the QW, the two orientations would have similar spacer layer dependence. Also, note that the sign of the out of plane polarization flips when the spacer layer $d$ is greater than 220 nm. This effect is also seen in the in-plane polarization, preserving the overall minus sign between the two orientations (not shown). Due to its spacer layer dependence, the sign flip for the d > 200nm devices suggest that its origin may be intrinsic to spin transport and is unlikely due to spin injection processes, however, further investigation is needed.

Further insight into the mechanism underlying the anisotropy is obtained by considering the background polarization from the EL. We plot the polarization data for all of the samples without the linear background subtracted to investigate the possibilities that the spacer layer dependence is due to modulation of the strain from the overlaying magnetic layer [Fig. 3(b)]. As mentioned earlier, the linear slope of the polarization's field dependence is due to Zeeman and strain contributions.[15] Clearly, the slope of the linear background is very similar for all the samples (even for d > 200 nm) and shows no systematic variation as function of spacer thickness, suggesting that the effects of strain are not the cause of the anisotropy. In addition, the non-magnetic reference sample has a different slope than the magnetic samples, supporting our assumption that the slope is sensitive to strain. Thus, the spacer layer dependence of the anisotropic spin injection efficiency and the sign flip at larger $d$ (> 200 nm) are not likely caused by strain variation in the sample set. While the mechanism is still unclear, we propose that this anisotropy could arise from either or the combination of the following: 1) anisotropy in the spin polarization of (Ga,Mn)As, 2) differing spin scattering mechanisms for HH vs. LH, or 3) spin scattering mechanisms that depend on spin orientation relative to the

transport direction.[16]

In summary, a fourteen-fold anisotropic electrical spin injection efficiency has been measured up to distances of 420 nm. EL polarization collected both parallel and perpendicular to the growth direction suggests that both heavy and light holes are spin polarized in the QW. Although the anisotropy mechanism is unclear, we are able to rule out effects intrinsic to the QW, such as optical selection rules and strain. The authors thank D. R. Schmidt and J. A. Gupta for technical support as well as P. A. Crowell, R.K. Kawakami and R. J Epstein for helpful discussions. Work supported at UCSB by the AFOSR F49620-99-1-0033, NSF DMR-0071888, and DARPA/ONR N00014-99-1-1096. The Japan Society for the Promotion of Science and the Ministry of Education in Japan support the work done at Tohoku University.

**Figure Captions**

Figure 1

(a) Spectrally-resolved EL intensity along the growth direction for several bias currents, I (note semi-log scale). Inset shows device schematic and EL collection geometries. (b) I-V characteristic. (c) EL intensity (solid curve) and polarization (●) at $H_\perp$ = 5 kOe showing a peak in the polarization at the QW ground state (E = 1.39 eV). (d) Magnetic characteristics of an unprocessed part of the sample when applying a field perpendicular (open squares) and parallel (solid curve) to the sample plane (note the different field scales).

Figure 2

(a) Temperature dependence of the relative changes in the energy-integrated [gray shaded area in Fig. 1(a)] polarization $\Delta P$ as a function of out of plane magnetic field. When T < 62 K, polarization saturates at $H_\perp$ ~ 2.5 kOe, commensurate with Fig. 1(d). Inset shows M(T), indicating that the polarization is proportional to magnetic moment. The absence of saturating polarization at T = 5 K from a (b) non-magnetic device and from a (c) magnetic structure under optical excitation. (d) Hysteretic EL polarization as a function of *in-plane* magnetic field reveals anisotropic spin injection efficiency giving rise to a magnitude difference and sign flip.

Figure 3

(a) Spacer layer dependence of EL polarization as function of out of plane field. Inset compares the *magnitude* of the polarization collected both in ($\Delta P_R$) and out of plane ($\Delta P_S$) as a function of spacer layer thickness. As *d* decreases $\Delta P_S$ monotonically increases from 0.5 to 7%, while $\Delta P_R$ remains unchanged. (b) All samples plotted without the background subtracted reveals Zeeman and strain related contributions. All magnetic samples have similar slopes suggesting spin injection anisotropy is not due to selection rule enhancement or strain.

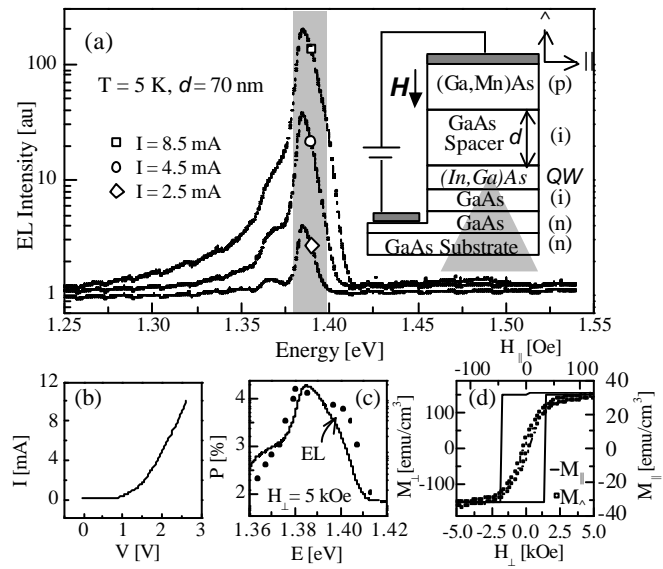

Figure 1. Young et. al.

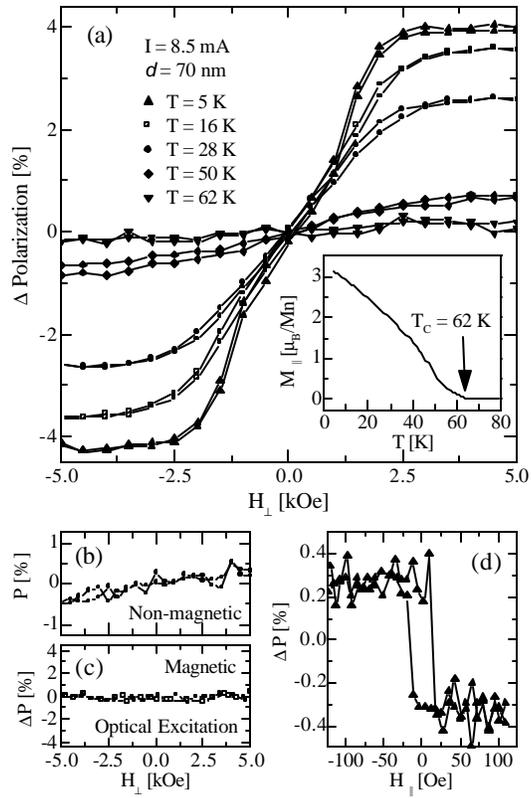

Figure 2. Young et. al.

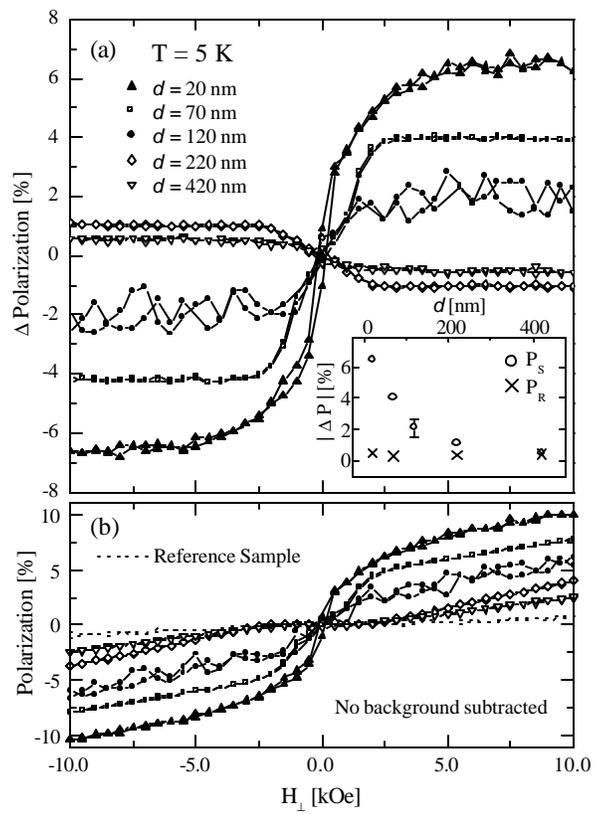

Figure 3. Young et. al.